# InCoRe - An Interactive Co-Regulation Model: Training Teacher Communication Skills in Demanding Classroom Situations

Chirag Bhuvaneshwara[1]
Lara Chehayeb[2]
Alexander Haberl[2]
Julius Siedentopf[2]
Patrick Gebhard[2]
Dimitra Tsovaltzi[2]

[1] K8 Institut für Strategische Ästhetik gGmbH, Germany
[2] DFKI GmbH, Kaiserslautern, Germany

**Abstract:** Socioemotional and regulation processes in learning are important. We add to the understanding of previous work on co-regulation processes in the learning sciences, extending the caregiver-child paradigm and focusing on the teacher-student relation by presenting an interactive co-regulation model and the methodology for developing empirically grounded systems for training teachers. We focus on the combination of classroom management and affect models and detail the use of a psychological model to operationalise and automate the interaction with the virtual student. We delve into an annotation scheme developed to capture teacher subjective psychological experiences during training and how these affect their co-regulation behavior with students and contributes to understanding the role of teacher emotional experiences and their consequences of co-regulation processes for classroom management. This research is also a contribution to developing hybrid AI systems.

## 1. Introduction

Teachers are often an important caregiver for kids and influence their socioemotional development. In analogy to parents, teachers who can reflect on and self-regulate their emotions, motives (emotional needs of oneself) and intentions (desired affect on the other) (Nišponská, 2023; Talia et al., 2021), and behaviours during conflict situations with students, can also reflect on the students' perspective (Anonymous), and help them regulate themselves (Nišponská, 2023). Understanding teachers' motives and intentions for their co-regulation behavior with the student is therefore important. Developing theory-based training systems with comparably affordable technology offers the most basic and sustainable solution to dealing with crises. We look at how teachers socialize students to deal with socioemotional classroom crises. We capture their subjective experiences and how these relate to their co-regulation behavior by integrating two psychological models: OPD - Operationalized Psychodynamic Diagnostics (OPD-3, 2023) and conflict resolution (Blake & Mouton, 1979) respectively. We implement an interactive co-regulation model: InCoRe.

Previous research in teacher education has concentrated on collaboration in teacher groups and shown the importance of socioemotional strategies in groups. Models exist to capture the socioemotional aspects of student discussions, and theoretical work provides a first approach to establish the relation between cognitive and socioemotional regulation. Recent research also investigates the interaction between teacher emotional conduct and classroom socioemotional experiences of students (Frenzel et al., 2021). Although considerable research exists on the role of emotion regulation with respect to individual characteristics that develop in early childhood and affect their response to co-regulation processes long term, only recently has there been an attempt to investigate the self- and co-regulating subjective socioemotional experiences and understanding motives and intentions thereof in the classroom consequences for learning (Nišponská, 2023), and successful programs for schools already exist (Fonagy et al., 2009). Nonetheless, training approaches to teachers' co-regulation skills are difficult to implement in teacher professional development, due to related costs and effort, and ethical dimension with regard to in-situ training. New technologies combining psychological and computational affect models in virtual reality enable developing research-based situative interactive training systems to support teacher co-regulation skills and to improve their communication skills. "Anonymous System" stands out among them as it focuses on training socio-emotional skills and classroom management, emphasizes the importance of respectful and professional interactions in a classroom and is based on models of subjective experiences. The effectiveness of the scenario afforded by the semi-automatic "Anonymous System" was evaluated positively (anonymous). Here, we focus on the InCoRe model and its use in modelling the teacher co-regulation process and explore:

**RQ1:** To what extent can the OPD model be applied to teacher co-regulation behaviors and identify motives?

**RQ2:** To what extent can the mapping between OPD conflicts and the corresponding motives and intentions be validated based on our data?

**Teacher communication skills, intentions and conflict regulation behavior:** Acquiring socio-emotional skills influences communication skills and professional behavior (Connolly et al., 2012). Capturing the subjective interpretations of the different parties in a co-regulation process is necessary to develop an interactive model and a training system. The most well-known is the Belief-Desire-Intention (BDI: Bratman, 1987) model, a cognitive architecture that explains how agents (humans or artificial) make decisions. Desires represent the agent's goals, and intentions are commitments to achieve them. The later EBDI (Pereira et al., 2006) model incorporated emotions to provide more realistic representations of human behavior, however, it does not address the fundamental cognitive approach of the model, as well as the fact that it addresses human perception from a top-down approach, to align with the needs of computational modeling. Moreover, these models do not capture the subjective experiences of agents. This is crucial, as individual's perception of their own intentions and conflict regulation styles could be different from their actual behavior (Holt & DeVore, 2005). Other approaches to computational models of affect (e.g. ALMA (Gebhard 2005)) are much more agile, with respect to capturing the emotional complexity, but do not offer a full modeling of the co-regulation process based on subjective emotional experiences and individual motives, and with regard to challenging situations. Conflict resolution strategies (CRS) need to be modelled, to be able to connect internal conflicts and co-regulation behavior. Internal conflicts influence interpersonal conflicts through processes of (negative) co-regulation. This is known as conflict regulation, negotiating perceived opposing motives, beliefs, values, or practices (Gelfand et. al., 2012) and can be interpersonal or intrapersonal.

Different degrees of affective empathy can be characterised through conflict resolution theory by following the dual concern, for self and others, model (Blake & Mouton, 1979) and can motivate conflict regulation strategies (CRS) along two dimensions, relationship and task: smoothing (high concern for others), forcing (high concern for self), compromising (medium concern for both), withdrawing (low concern for both ); problem-solving (high concern for both). High arousal levels, like in conflict situations, may reduce the ability of empathic response and perspective taking and trigger the switch of emotion regulation to automatic response through fight or flight. This switch may be influenced by different factors such as internal conflicts typical for a person (Gross, 2015). Perspective taking and self-awareness are directly influenced by one's internal conflict. We use the model of conflict resolution strategies to translate the teacher's OPD motives into behavioral intentions based on the conflict resolution strategies (Sirois, 2015). Thus, InCoRe extends previous models.

**Conflicts in the OPD Model:** The OPD-Model was developed for therapeutic observation and diagnostics. It aims to analyse subjective human emotional experiences bottom-up, focusing on observable behaviors and their underlying motivations and conflicts, which humans may not always be aware of. Since it takes the stance of deviations from productive human behavior, it can be applied to address subjectively experience socioemotional challenges. It also offers a very situative focus, so that it can be adjusted and applied for different interaction scenarios. Internal conflicts stem from experiences during development and are triggered by present interactions. Internal conflicts are experienced when one's desires or socio-emotional needs are not met. The dominant emotional state, the lead affect, influences how one behaves to reach the desired state in the interaction (Gross, 2002). Furthermore, the behavior is influenced by the teacher's intention, which is shaped by generally accepted social norms. (Fonagy et al., 2006). In our training, we aim to track the lead affect of the teacher during the automated interaction, based on the OPD method (Group O.W. 2001). In the Wizard of Oz (WoZ) experiment, a trained interviewer conducted an interview with the participant to reflect on their experience in the VR classroom. The aim is to understand the participant's intentions behind specific behaviors and to identify the lead affect driving their actions. We analyze the connection between the self-reported internal conflict of the teacher, the lead affect and the teacher's intentions. In the final automated system, we utilize this data to model the automatic responses of the virtual student agent, as well as the co-regulating feedback agent designed to train self-compassion. InCoRe is a first attempt to implement an interactive model of co-regulation based on the OPD model, while using advantages of previous computational models. It integrated the ALMA model. However, the OPD does not offer a clear implementable model on how to capture the individual interpretations of a given human interaction situation. Social norms have been used to model situation-based dialogue interactions to capture obligations that are not captured otherwise. InCoRe uses social norms to evaluate student behavior in each turn from the teacher perspective.

**Social Norms and Obligations:** In our Automated System, a Social Norm represents a set of classroom rules for the teacher and for the student that become active in a specific context. Each social norm contains a list of social obligations, and the active social norms are identified by the discourse analyser component in Geni:OS (García Ucharima, R. 2023). In the context of dialogue management for tutorial dialogues, social obligations have been used to model and automate feedback using the general framework of dialogue management and dialogue acts, which focus on language communication (Anonymous) This modelling considers social roles, the history of interaction, and differentiates between the task (tutoring) and the socioemotional interaction level. InCoRe uses the social norms and obligations paradigm, to represent subjective interpretations of the situation based on the perceived social role of oneself and of the collocutor.

**VR-Systems as Training Environments:** Research in virtual 3D learning and simulation environments indicates significant potential for enhancing learning and training outcomes through these technologies. In the domain of teacher education, mixed reality (MR) applications provide an opportunity for aspiring educators to practice strategic decision-making in controlled and safe settings (Dalinger et al., 2020). These environments, incorporating virtual interactive agents, have demonstrated substantial benefits for fostering social and emotional learning. Existing MR-based conflict training systems often rely on scripted scenarios with pre-recorded student reactions performed by actors (Dalinger et al., 2020). While these systems provide authentic experiences, they may fall short in offering tailored training options. Understanding one's emotional states and needs is pivotal in such contexts. Complementary tools, such as group awareness technologies (GATs), combined with AI-based behavioral support, have shown promise in enhancing communication, perspective-taking, and group interactions in learning environments (Anonymous).

Although numerous projects aim to create secure learning environments, their primary focus has often been on enhancing knowledge acquisition and pedagogical strategies. Our training distinguishes itself by prioritizing the development of socio-emotional competencies and effective classroom management skills. Central to our training system is the emphasis on fostering respectful, professional classroom interactions as a foundational element, independent of specific teaching methods or content. Moreover, the system addresses supporting teacher well-being, highlighting its role in sustaining effective educational practices.

**Our Training System:** is a virtually immersive collaborative 3D environment that analyzes social signals and synthesizes apt virtual character behaviors to train teachers for effective conflict regulation behaviors in the classroom. Our system is implemented in two versions: 1) Wizard of Oz (WoZ) System and 2) Automated System. The WoZ system collected data to validate separate aspects of behavior modeling, and the collected data is used for designing InCoRe approach and also to develop machine learning models in the Automated System. In our training, based on how the teacher addresses the conflict with the virtual student, the InCoRe Model utilizes the concepts from OPD to generate apt student behavior using Visual SceneMaker (VSM). In the case of the WoZ system (anonymous), there is a human wizard in the loop, who evaluates the teacher's behavior in real time and supplies the emotion understanding that is needed to ascertain the teacher's intention for appropriate student behavior generation. However, for the automated system (anonymous), InCoRe, without any human intervention, performs emotion understanding and derives teacher intentions, in real time, using OPD. In this paper, we present the theoretical underpinnings of the InCoRe model and its implementation for our system to train teachers' socioemotional behavior. We report on results from an empirical study on the main theoretical assumption of the model, the connection between subjective socioemotional experiences and co-regulation behaviors of teacher trainees, which informed the final implementation of InCoRe, and on some indicative results about the final system.

## 2. Theoretical Underpinnings of the InCoRe Model

To model the socio-emotional aspects of co-regulation we combine the OPD model, which describes internal emotional experiences that depend on the situation. We use social norms to model how a person experiences themselves in a specific situation through their social role in the situation. Specifically for the teacher-student co-regulation, and to define the communication skill of the teacher, the social norms provide the basis to capture the internal emotional experience. This modeling, connecting OPD and social norms, lies at the heart of the co-regulation model. It can represent that the better the teacher reflects and regulates their own emotional reaction, the better they can react empathically, verbally and non-verbally, and corregulate the student. We define this as reflected empathic communication, where the collocutor listens and is empathically „attuned“ to the affective experience and needs of the other person, depending on their social role (Fonagy et al., 2009). For example, if a teacher tends to overvalue classroom rules, they may feel threatened by the student disregarding them, e.g. using a mobile phone, and force the student to comply with them (Figure 1).

Using the OPD, we can match the teacher's emotional experiences and analyse their reactions to the student. InCoRe follows three models in each turn to evaluate the teacher's co-regulation behavior and decide on the student response, here implementing a common OPD conflict.

1. **Connecting Lead Affect and OPD internal Conflict:** We model the current OPD internal conflict based on the multimodal modelling of emotion providing the lead affect and the ALMA.
2. **Connecting social-obligations & OPD internal Conflict**: We recognise the social obligations identified in each turn from our situative (here classroom) social norms and obligations taxonomy. Given the teacher conflict we can decide on the most important obligation of the student for the teacher, and on the teacher intention thereof, and rank the list of social obligations in each term for the subjective importance to the trainee, as a characteristic of the teacher's social interaction structure.
3. **Connecting Conflict and CRS:** Each OPD conflict is associated with certain motives, what the teacher aims to achieve in social interaction. Given the OPD conflict the teacher's intentions are characterised for the conflict resolution style they represent.

## 3. InCoRe Architecture Embedded in our Automated Training System

The automated system uses Geni:OS[1] and AffectToolbox (Martes et al. 2024) to perform automated social signal analysis, which is then processed further by the InCoRe Model (Figure 1). InCoRe first models teacher co-regulation behavior to drive the student co-regulation behavior and generate appropriate student reactions. VSM implements InCoRe in a *Feedback* and *Control Loop*, which synchronizes all the software components, uses InCoRe to further process the social signals and appropriately controls the student agents.

**Figure 1**

*InCoRe architecture showing the different software components interacting within a dialogue turn.*

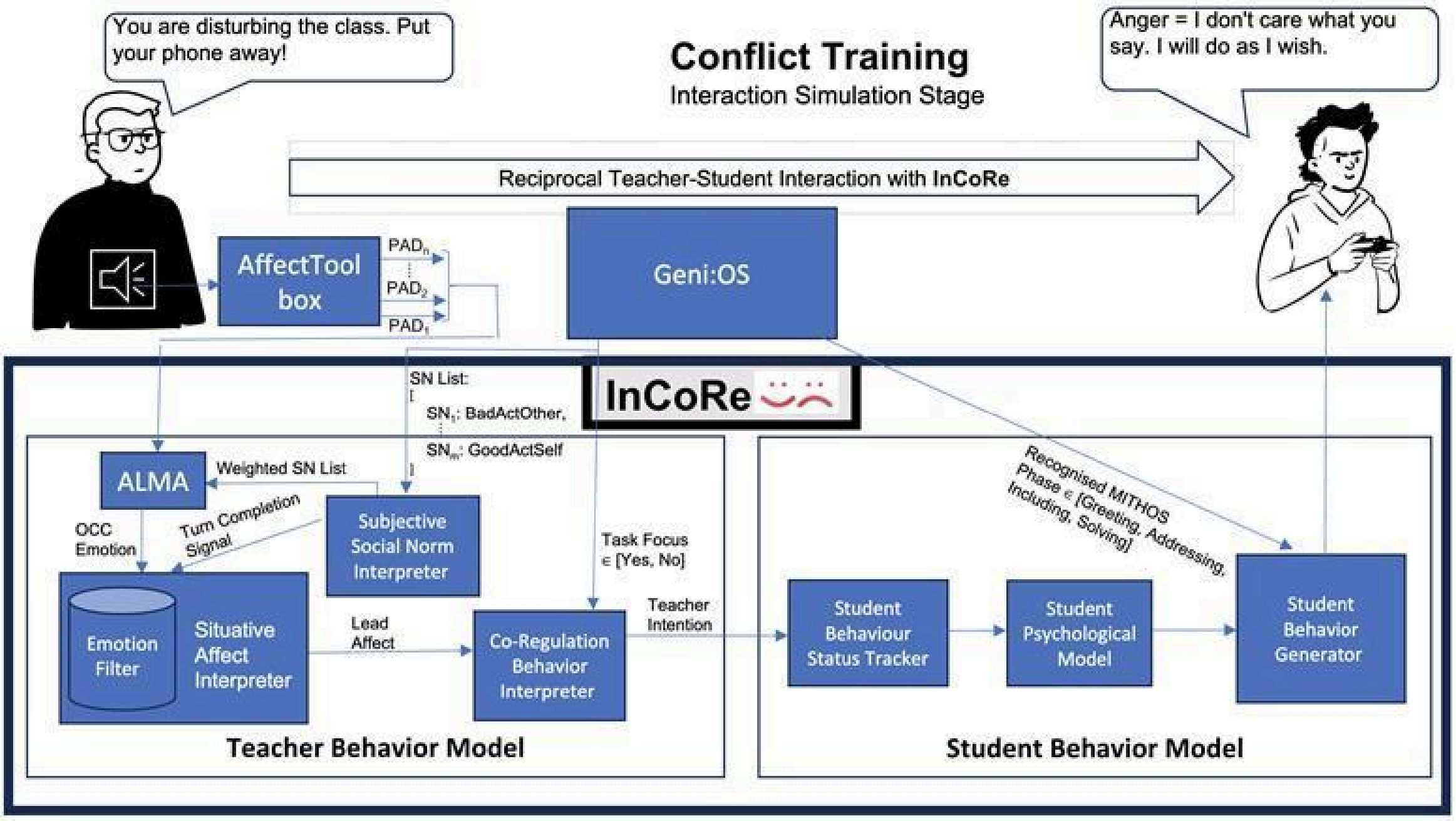

### 3.1 Teacher Co-Regulation Behavior Model

This model predicts the teacher's intention, how the relationship between teacher and student evolves and keeps track of what progress the teacher made in the task. This is done in a turn-based manner after each student-teacher interaction. An interaction is modeled as the end of a conversational turn where the student does or says something and the teacher reacts verbally. The model consists of several components (see Figure 1): **1) ALMA:** is used to convert the social signal analysis results into OCC Emotion. **2) Social Norm Interpreter:** Geni:OS sends a list of followed/broken social norms to InCoRe. They are processed in the Subjective Social Norm Interpreter, which weighs the list of social norms according to the current conflict type. **3) Situative**

[1] https://www.semvox.de/technologies/Geni:OS/

**Affect Interpreter:** is triggered by the end of the turn and it processes the logged emotions of the current turn to determine the current Lead Affect. The fist step for this is to apply the **Emotion Filter** to determine the strongest emotion(s) of this turn, which is then mapped to the Lead Affect. **4) Co-Regulation Behavior Interpreter:** determines the teacher's intention, represented by the conflict resolution style, by mapping the Lead Affect to a change in relationship level. It uses the task focus provided by Geni:OS and determines the conflict resolution style (anonymous, anonymous).

### 3.2 Student Co-Regulation Behavior Model

This model keeps track of the relationship and task levels and generates student behavior as described in (anonymous, anonymous). The **Student Behavior Status Tracker** keeps track of these two variables, and uses them to drive student behavior in the conflict situation with the teacher. The **Student Psychological Model** refers to the personality of the student instigating the teacher (anonymous, anonymous). **Student Behavior Generator** is a set of conditional animation commands in VSM to control the virtual student in Tricat classroom (anonymous). Thus, the reciprocal reaction of the student (based on the implemented internal conflict) on the other end is a reaction based on the OPD model, which starts the new turn.

## 4. Empirical Evaluation of InCoRe – OPD Conflict and Motives

Participants were recruited from higher semester students in teacher training programs at universities. A total of 60 German-speaking students participated. Ethical approval was obtained appropriately. Participants engaged in a simulated classroom scenario, interacting with virtual students. A semi-structured interview was conducted prompting participants to reflect on specific moments of the interaction by watching video recordings and through detailed questions about their subjective emotional experience of the conflict, their motives and intentions. Additionally, participants completed the OPD lead affect questionnaire to assess their emotional reaction, and the interviewer completed the OPD countertransference questionnaire, to cross-reference the emotional reaction of the participants (OPD-2, OPD-3).

To systematically capture the types of conflicts in teacher-student interactions within our system and empirically validate their use for the final automatic InCoRe in inferring teacher motives, an annotation scheme was developed (Table 2) based on the OPD model (OPD-2; OPD-3). This scheme categorizes conflicts by thematic types of internal conflicts, which refer to psychological struggles within an individual related to personal needs, desires, and self-perceptions. These conflicts often involve opposing motivations or feelings, such as a desire for control versus submission in the "Submission vs. Control" conflict implemented here for the student and hence elicited by the teacher, and each conflict is assessed for modulation as either active or passive (Table 2). This approach enables a nuanced analysis of socio-emotional dynamics in teacher training, identifying modelling the most often identified both conflict types and their modes for a user based on the current situation, within classroom scenarios. This is an attempt to capture the teacher's behavior, bridging the gap between an individual's perception of their own intentions and conflict regulation styles that could be different from their actual behavior (Holt & DeVore, 2005). Such measurement assumes less social desirability and perceived realism of hypothetical situations.

The annotation scheme was developed following the process of Mayring (2014). The coding rules were developed based on the OPD theory on conflicts, and how these surface in the interaction and in the self-reports about motives (OPD-2; OPD-3). The coding rules were in general developed with a focus on their 1) theoretical integrity; 2) ease to apply by independent annotators 3) generalisability. Two annotators, a research assistant and an experienced expert psychodynamic therapist, worked together following multiple iterations of development and consultation to reach a consensus, integrating both the OPD theory and clinical expertise. Together, they coded a small sample of the transcribed interviews (2-9, 12,5% of the total 56 interviews), for teacher motives, following also the Apprenticeship Model (Renkl, 2001), which emphasizes training through close collaboration. This model supports the reliability of our process, focusing on in-depth training and agreement before independent coding. The framework method (Gale et al., 2013) was employed to systematically organize the data. The coding rules were adapted making sure that the identified teacher motive descriptions from the data also matched the OPD description of the motives associated with each conflict type (theoretical integrity). This is relevant in our system, since we base our interpretation of motives and the identified conflict. We then formulated the coding rules in a generally understandable not too strict manner (generalisability), as teachers are unlikely to use the psychotherapeutic vernacular, and to avoid having a too large fitting (being too generous in identifying conflicts). Our rationale was that the OPD model might alternatively be too strict, as it was mainly designed for pathological cases of internal conflicts. Once a consensus was reached between the annotators, anchor examples were developed that apply to our data at best for independent annotators (ease of application).

We tested the theoretical relation between the OPD internal conflict, and the teacher motives, which we use to automatically model the co-regulation behavior of the teacher. We formulate the following hypotheses. With regard to applying the OPD model for teacher co-regulation behavior (RQ1), since the OPD model provides a mapping between the motives, our hypotheses are:

**H1a:** Using the InCoRe annotation scheme, the motives and conflicts theoretically described in the OPD model can be identified in the collected data.

**H1b:** The conflict experienced by the participant, as derived from interview annotations (objective data), correlates with the OPD conflict implemented in the student agent.

With regard to model validation (RQ2), we test:

**H2a:** The self-reported OPD conflicts correlate with the annotated OPD conflicts of observed motives.

We further analyzed which conflicts and whether their passive or active modi are more prominent in the teaching context. This has implications on the focus of the implementations, which conflicts are worth implementing, but also for the validation of the training approach, as it is expensive to implement all conflicts on the one hand, but still useful, if teachers experience them. Finally, we explored the alignment between the annotated conflicts from the WoZ-Study, and the detected conflicts using automatic InCoRe a later study, where users train in the fully interactive system and the conflict data come from the interactive phase, without a reflective intervention as in the interview. We expected to find at least main prompted conflict ("control").

**Table 1:**
*Abbreviated example of coding rules for the OPD motives and matchings conflicts*

| Conflict | Value | Definition | OPD Example | Code Rule | InCoRe Application |
|---|---|---|---|---|---|
| K2: Submissiveness vs. Control (active) | Exercising control over yourself and others. | "..aggressively striving for dominance.., ..to never give up control in order to exclude further helplessness. [..] If the conflict escalates, it is about sheer power" (OPD-3, p.128) | "The main thing is that I have the upper hand!" (OPD3) | Subject wants control or influence on the behavior of others, e.g. class events and the behavior of students or on processes... | Teacher leaves little room for maneuver, sets a timetable for everything, and checks meticulously whether instructions are followed. Reacts very sensitively to self-confident students |

The data comprises interviews conducted with 24 participants, each interview (≤ 80 minutes) captures teachers' reflections on socio-emotional challenges and conflicts in our training, providing a rich foundation for qualitative and quantitative analysis. To ensure scientific rigor, an independent coder applied the developed coding, adopting a reflexive approach with detailed notes to address potential biases and align with the theoretical framework (Finlay, 2002). The sample of 24 participants was randomly selected from an initial pool of 57, starting from the first interviews that were transcribed and made available to the independent annotator. This approach ensured a degree of randomness while also leveraging the earliest data, providing a diverse experience range. The combined approach of collaborative development, reflexivity, and structured analysis enhances the study's robustness, offering a solid foundation for analyzing conflict types and teacher-student interaction dynamics. A total of 250 coded conflict instances were annotated to identify conflict distribution patterns, frequency of active vs. passive modes, and the emotional dynamics associated with each conflict.

RQ1- H1a and H1b: We used the annotations of the motives and conflicts evident in participant interactions during the interview. We analyzed the annotated conflicts to determine whether the experienced conflicts reported by participants could predict the corresponding OPD conflicts implemented in the student agent. This analysis aimed to validate the theoretical constructs of the OPD model within our specific educational context. RQ2 - H2a and H2b: We analyzed the relations between participants' self-reported conflicts, the interviewer's countertransference, and the annotated conflicts derived from the interviews, to validate the mapping between OPD conflicts and the corresponding motives.

**Results of lead affects and countertransference:** To compare annotated conflicts and reported reactions (lead affects & countertransference) and to take into consideration the complexity of emotional experiences, each coded emotion was entered into subcategories. Subcategories A_1, A_2, A_3, and A_4 were defined to account for cases where participants provided multiple equally strong responses to a conflict situation. The first occurring response was assigned to A_1, the second to A_2, and so forth. For participants who reported only one dominant response, this was consistently included across all subcategories. We analyzed the data using Chi-square to examine relations of nominal data and Cramer's V for the effect size. The analysis of the lead affects revealed consistently significant relations between participants' reported reactions and the annotated conflicts. The Chi-square tests for subcategories A_1 to A_4 were significant (A_1: $x^2(70)=108.96$, $p=0.002$, $V=0.58$, $N=47$); A_2: $x^2(70)=96.47$, $p=0.02$, $V=0.54$, $N=47$; A_3: $x^2(70)=106.21$, $p=0.003$, $V=0.568$, $N=47$; A_4: $x^2(70)=107.39$, $p=0.003$, $V=0.57$, $N=47$), indicating moderate to strong relations between lead affects and the annotated conflicts. The analysis of countertransferences yielded mixed results. A strong significant relation was found in Category A_2 ($x^2(70)=109.20$, $p=0.002$, $V=0.57$, $N=47$). The other results were not significant (A_1: $x^2(49)=56.31$, $p=0.22$, $V=0.41$, $N=47$; A_3: $x^2(56)=69.41$, $p=0.11$, $V=0.46$, $N=47$; A_4: $x^2(49)=61.64$, $p=0.11$, $V=0.43$, $N=47$), but showed moderate relations.

**Figure 3a:**
*Distribution of Conflict Types*

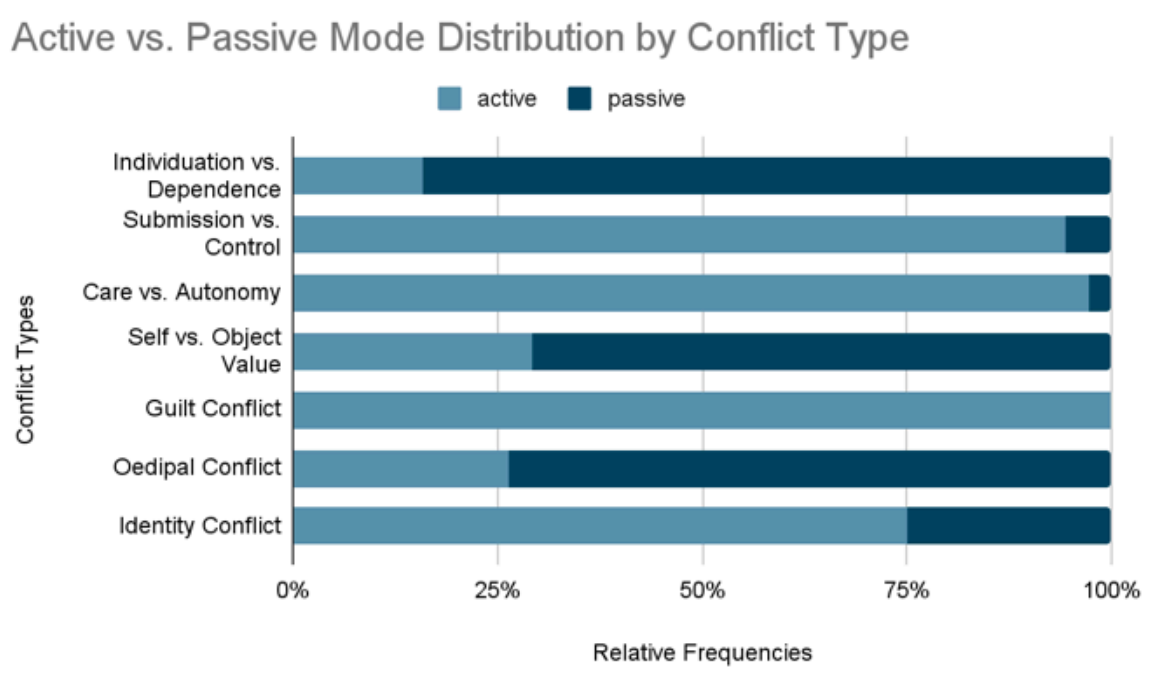

**Figure 3b:**
*Active/Passive Mode Distribution (Table 2, confl. ref)*

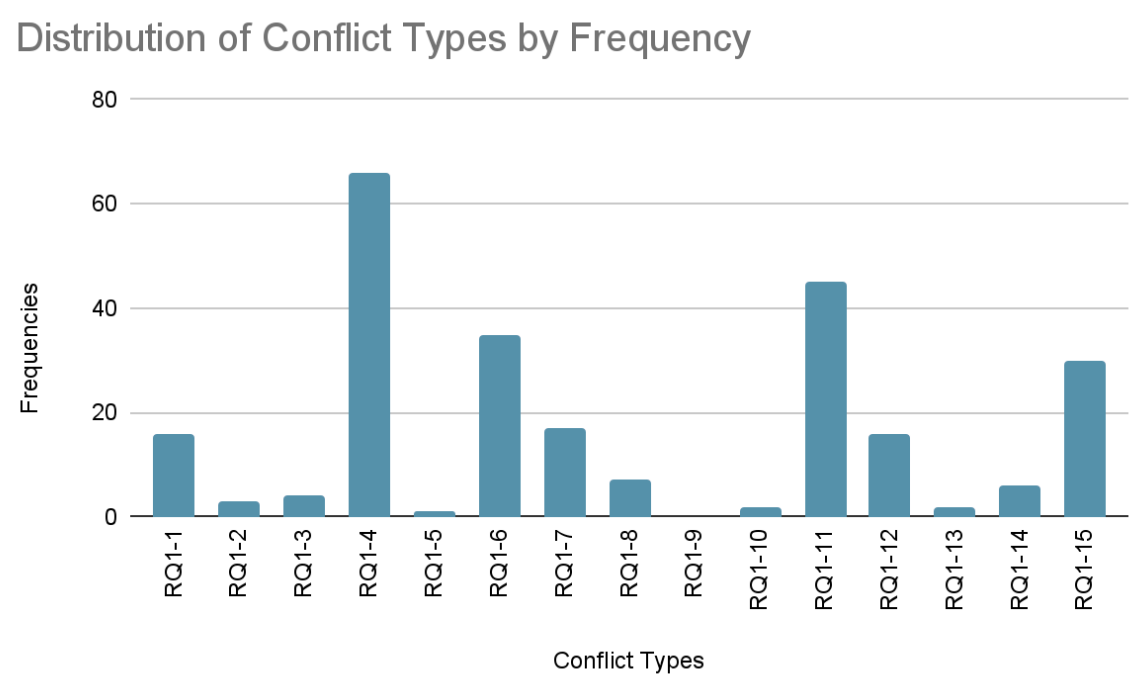

**Table 2:**
*Summary of most frequent conflict-type with reference key and results (right) from the interview annotation with coloring for the most frequent ones - deeper grey: the most frequent one)*

| Confl. Ref. | Conflict Type - Mode | Frequencies |
|---|---|---|
| RQ1-1 | K1: Autonomy vs. Dependence - Passive | 16 (6,4%) |
| RQ1-4 | K2: Submission vs. Control - Active | 66 (26,4%) |
| RQ1-5 | K3: Need for Care vs.Autarky - Passive | 1 (0,4%) |
| RQ1-6 | K3: Need for Care vs. Self-Autarky - Active | 35 (14%) |
| RQ1-7 | K4: Self-Worth Conflict (Self vs. Object Value) - Passive | 17 (6,8%) |
| RQ1-11 | K6: Oedipal Conflict - Passive | 45 (18%) |
| RQ1-12 | K6: Oedipal Conflict - Active | 16 (6,4%) |
| RQ1-15 | K0: Conflict Defense | 30 (12%) |

**Analysis of Conflict Types:** The analysis reveals that "Submission vs. Control" emerges as the most prevalent, particularly in active mode “Control”, accounting for 66 instances or 26.4% of all conflicts. This high frequency underscores a significant emphasis on authority and discipline in teacher-student interactions. This result validates the use of the model for the interactive Training, as it is a direct response of the participant to the same conflict implemented for the student agent. Moreover, the "Oedipal Conflict," markedly present in the passive mode with 45 instances (18%), highlights the internal determination to negate any emotions that may be sexually connoted as quite high among teacher trainees, who prefer to stay in the background rather than show shame. Interestingly, the conflict "Need for Care vs. Autarky" demonstrates a particularly proactive engagement, with 35 instances (14%) compared to just 1 in passive mode (0.4%). This indicates that teachers handle autonomy issues assertively. In contrast, conflicts such as "Self-Worth Conflict" are predominantly expressed in passive modes 17 instances (6.8%), pointing to a more reflective, but also self-blame approach.

The main results of the distribution of conflicts in active and passive modes (Table 2) showed a difference for passive "Autonomy vs. Dependence" (16 vs. 3), which suggests avoiding direct confrontation, through an excessive fear of losing contact to the student, and for active "Need for Care vs. Autarky" (34 vs. 1), indicating a direct offensive approach to autonomy-related conflicts, likely due to the authoritative role of teachers.

**Comparative results annotations and automatic InCoRe:** To further verify if the implementation of InCoRe captures subjective co-regulation behavior in its fully automated version, we analysed the detected lead affect in a new still running study. After each interaction turn, we recorded the analysed lead affect and corresponding conflict. We analyzed the data from 20 Participants (total of 275 turns and lead affects). We computed the frequency of occurrence of all lead affects across all participants. The data reveal that the most common detected internal conflict aligns in the automated InCoRe with the intended conflict: Control - submissiveness active 55.63% vs. 26.4% annotated (defiant aggressiveness and wanting the complete control), and also points to conflict denial 13.82% vs. 12 % (acting non-emotional and not admitting to the students' conflictual behavior) as a prominent conflict, aligning with the results from the annotated conflicts.

## 5. Discussion

The study explored the applicability of the OPD model for analyzing co-regulation in teacher-student interactions. These results provide evidence for the OPD model's potential to capture emotional dynamics in educational contexts and emphasize the importance of addressing underlying psychological mechanisms in co-regulation training. In summary, the results demonstrate that lead affects consistently exhibit significant and moderate to strong relations with the annotated conflicts, unlike countertransferences, where only one subcategory showed a significant relation, but all all were moderate effects. The findings show a high prevalence of the "Submission vs. Control" conflict active mode, indicating a pronounced tendency among teacher trainees to exercise authority and maintain disciplinary control. The analysis of lead affects and countertransferences revealed significant correlations between reported reactions and annotated conflicts, especially for lead affects. However, countertransferences showed weaker and less consistent correlations, underscoring the difficulty of applying the countertransference method for non-expert therapists. Discrepancies between situational reactions (questionnaires) and interview annotations, particularly in self-worth conflicts, suggest that immediate emotional defenses may shift toward more reflective processing over time. The comparative results of the InCoRe theoretical assumptions and the automatic version show a tendency to align for the strongest prompted response, control (through the student internal conflict), but interestingly also for conflict avoidance, probably due to teachers disregarding socioemotional conflicts as unimportant. These findings validate the scenario, InCoRe and stress the necessity of co-regulation training for teacher trainees.

Overall, the study contributes to understanding teachers' emotional experiences and demonstrates the InCoRe model's power to capture nuanced co-regulation processes for teacher education. Following recent research which investigates the self- and co-regulating subjective socioemotional experiences and understanding motives and intentions in the classroom for learning (Nišponská, 2023; Talia et al., 2021), and relevant (Fonagy et al., 2009), here we have shown that a systematization of the co-regulation behavior is possible by extending existing model of subjective psychological experience and conflict regulation.

## 6. Conclusion and Future Work

In this paper we presented an interactive model of co-regulation, InCoRe which manages reciprocal personalized responses of a virtual student. We presented an approach and its validation, using psychological models, analysed and discussed teacher trainee conflicts based on their motives, which defines their co-regulation strategies in the classroom. The work presented contributes to understanding the role of teacher subjective emotional experiences and their consequences for classroom management. It adds to understanding of co-regulation processes focusing on the teacher-student relation and extending the caregiver-child paradigm. It is a contribution to developing hybrid AI systems for learning sciences. More testing and more implementations of conflicts are needed.

## References

Blake, R. R., & Mouton, J. S. (1979). Intergroup problem solving in organizations: From theory to practice. In W. G. Austin & S. Worchel (Eds.), The social psychology of intergroup relations (pp. 19–32). Monterey, CA: Brooks/Cole.

Bratman, M. E. (1987). Intentions, plans, and practical reason. Harvard University Press.

Connolly, T. M., Boyle, E. A., MacArthur, E., Hainey, T., & Boyle, J. M. (2012). A systematic literature review of empirical evidence on computer games and serious games. Computers & Education, 59(2), 661–686.

Dalinger, T., Thomas, K. B., Stansberry, S., & Xiu, Y. (2020). A mixed reality simulation offers strategic practice for pre-service teachers. Computers & Education, 144, 103696. JMIR Serious Games. (2020). Volume 8(3), e18822

Fonagy, P., & Bateman, A. W. (2006). Mechanisms of change in mentalization-based treatment of BPD. Journal of Clinical Psychology, 62(4), 411–430.

Fonagy, P., Twemlow, S. W., Vernberg, E. M., Nelson, J. M., Dill, E. J., Little, T. D., & Sargent, J. A. (2009). A cluster randomized controlled trial of child-focused psychiatric consultation and a school systems-focused intervention to reduce aggression. Journal of Child Psychology & Psychiatry, 50(5), 607–616.

Gale, N. K., Heath, G., Cameron, E., Rashid, S., & Redwood, S. (2013). Using the framework method for the analysis of qualitative data in multi-disciplinary health research. *BMC Medical Research Methodology, 13(1)*.

García Ucharima, R. (2023). A computational model for discourse analysis, applying knowledge about social norms & emotions (Master's thesis). Retrieved from Saarland University Repository.

Gelfand, M. J., Leslie, L. M., Keller, K., & de Dreu, C. (2012). Conflict cultures inorganizations: How leaders shape conflict cultures and their organizational-level consequences. Journal of Applied Psychology, 97(6), 1131.

Gross, J. J. (2002). Emotion regulation: Affective, cognitive, and social consequences. Psychophysiology, 39(3), 281–291.

Gross, J. J. (2015). Emotion regulation: Current status and future prospects. Psychological Inquiry, 26(1), 1–26.

Group O.W. and A. (2001). Operationalized psychodynamic diagnostics: Foundations and manual. Seattle; Toronto: Hogrefe & Huber.

Holt, J. L., & DeVore, C. J. (2005). Culture, gender, organizational role, and styles of conflict resolution: A meta-analysis. International Journal of Intercultural Relations, 29(2), 165–196.

Mayring, P. (2014). Qualitative content analysis: Theoretical foundation, basic procedures, and software solution (p. 145).

Nišponská, M. (2023). Manipulation of Time Continuity in Shared Narratives. On the Construction of Collective" Truths" and Its Ambivalent Function in the Social World and in Education. Historia Scholastica, 9(2).

Park, S., & Tsovaltzi, D. (2022). Implicit and explicit emotion regulation for conflict resolution: Narrative and self-compassion as anti-bullying training. In Proceedings of the 16th International Conference of the Learning Sciences (ICLS 2022) (pp. 187–194). International Society of the Learning Sciences.

Pereira, D., Moreira, N., & Oliveira, E. (2006). Modelling emotional BDI agents. Retrieved from

Renkl, A. (2001). Situated learning: out of school and in the classroom. In *Elsevier eBooks* (pp. 14133–14137).

Sirois, F. M. (2015). A self-regulation resource model of self-compassion and health behavior intentions in emerging adults. Preventive Medicine Reports, 2, 218–222.

Talia, A., Duschinsky, R., Mazzarella, D., Hauschild, S., & Taubner, S. (2021). Epistemic trust and the emergence of conduct problems: Aggression in the service of communication. Frontiers in Psychiatry, 12, 710011.

Tsovaltzi, D., Judele, R., Puhl, T., & Weinberger, A. (2017). Leveraging social networking sites for knowledge co-construction: Positive effects of argumentation structure, but premature knowledge consolidation after individual preparation. Learning and Instruction, 52, 161–179.

## Acknowledgments

This study is funded by the XXXXXXX within the funding line "Interactive systems in virtual and real spaces - Innovative technologies for the digital society" (Project XXXXXXX, grant XXXXXXX).